\documentstyle[pra,aps,multicol,amssymb,epsfig]{revtex}
\begin{document}

\title{Discrete instability in  nonlinear lattices}
\author{ J. Leon, M. Manna\\
{\em Physique Math\'ematique et Th\'eorique, CNRS-UMR5825,}\\
 Universit\'e Montpellier 2, 34095 MONTPELLIER (France)}
\maketitle

\begin{abstract} The discrete multiscale analysis for boundary value problems
in nonlinear discrete systems leads to a first order, strictly discrete,
modulational instability  (disappearing in the continuous envelope limit)
above a threshold amplitude for wave numbers beyond the zero of group velocity
dispersion.  Applied to the electrical lattice  [Phys. Rev. E, {\bf 51}, 6127
(1995)], this acurately explains the experimental instability at wave numbers
beyond $1.25$ rad.cell$^{-1}$. The theory is also briefly discussed for the
sine-Gordon and Toda lattices.  \end{abstract} 

\begin{multicols}{2}

\paragraph*{Introduction.}

The  Benjamin-Feir instability describes the mechanism of the 
exponential growth of the modulation of a wave  evolving in a continuous 
nonlinear dispersive medium \cite{benj-feir,bespalov}. It has revealed as a
fundamental process occuring in various physical situations, from water waves
to nonlinear optics through plasma waves, electrical transmission lines,
etc..., see e.g.  \cite{remsnet}.  
This instability, originally established for periodic
wave trains (Stokes waves in surface water theory) \cite{benjamin},
is actually a generic process in physics, well understood in the context
of the nonlinear Schr\"odinger equation (NLS) \cite{stuart} \cite{akhmediev}.
In the case on an infinite medium, and for the focusing NLS, this instability
has been shown to  evolve eventually into localized coherent nonlinear
structures (the solitons), see e.g. \cite{wadati}.

The analysis  of a nonlinear system is performed by first deriving from the
original model a simpler limit equation which usually results to be the
NLS. This universal limit is obtained by a very general approach called 
multiscale perturbation analysis \cite{taniuti}. 
It is worth remarking already that in such cases the instability is actually a
second order effect: NLS is obtained as a first order perturbation and the
Benjamin-Feir instability results from a perturbative analysis of NLS itself.

In the case of discrete systems where {\em intrinsic nonlinear modes} (called
also discrete breathers) are known to exist \cite{mackay}-\cite{karl}, the
modulational instability, mechanism for their spontaneous generation out of an
initial disturbance, is usualy described in a semi-discrete multiscale
analysis where the envelope of the discrete carrier wave is treated as a
continuous function \cite{semi-dis}-\cite{remsnet-elec-cont}. Fully discrete
treatments have also been proposed as in \cite{remsnet-elec-dis,kiv-peyr},
based on the {\em rotating wave approximation} which essentially
consists in neglecting harmonics. In a nonlinear system, this procedure
evidently may give wrong predictions for large times.

In some of these studies however, the problem is not that of the long time
evolution of an initial perturbation but rather the far behavior of an input
boundary disturbance (or forcing). This is particularly evident in the
electrical transmission line of \cite{remsnet-elec-cont,remsnet-elec-dis}
where an input signal (modulated wave) is sent at one end of the lattice
and collected at various locations along the line.

Hence we consider here the problem of the scattering of a plane wave incoming
into a nonlinear discrete medium. Based on the method founded in \cite{gros}, a
discrete change of variables is introduced and used to construct the related
discrete limit model. This results as a NLS-like equation in discrete space and
with space and time exchanged. For wave numbers beyond the zero of the group
velocity dispersion, it is shown that the waves are uniformly unstable as soon
as their amplitudes exceed a threshold value. Moreover this instability is a
purely discrete property as it is shown to disappear if the envelope is
treated as a continuous function like in the semi-discrete approach.

This approach is applied to the electrical lattice of \cite{remsnet-elec-dis}
where the problem is indeed that of the far behavior of an input boundary datum
and for which a set of very detailed and acurate experimental results are
displayed. In particular the measured threshold of modulational instability
occurs at wave numbers beyond $1.25$  (in rad.cell$^{-1}$) which is the value
predicted by the present theory. We predict also a threshold amplitude
depending on the input carrier frequency and modulation. For one of the
experiments of \cite{remsnet-elec-dis} this threshold value is $0.23\ V$,
significantly smaller than the amplitudes actually used in the experiments
where damping had to be overcome. This threshold amplitude is shown to
depend on the relative frequencies of the carrier v.s. modulation and hence
can be easily varied which should motivate new experiments.

\paragraph*{Basic tools.} 

Treating a boundary value problem corresponds to making an expansion of the
wave number around small deviations of the frequency from the linear dispersion
law.  To see this, we consider a plane wave 
\begin{equation}A(n,t)=\exp[i(Kn-\Omega t)] \end{equation}
in a nonlinear lattice (with intersite spacing normalized to 1) 
with the linear dispersion relation of bandpass filter type
\begin{equation}\label{lin-disp}
\Omega^2=\omega_0^2+4u_0^2\sin^2(K/2)\ .
\end{equation}
Due to nonlinearity, the actual frequency $\omega$ and wave number $k$
experience  deviations form the values $\Omega$ and $K$ obeying
the linear dispersion relation  (\ref{lin-disp}).
There are two ways of representing these deviations. 

With the wave packet
\begin{equation}
u_{n}(t)=\int dk\ \hat u(k) e^{i(kn-\omega t)},\end{equation}
we may expand the  frequency in terms of the assumed small deviations from 
the value $K$  of the wave number as
\begin{equation}
k=K+\epsilon\lambda\ ,\quad
\omega=\Omega+\epsilon v_g\ \lambda +
\epsilon^2 P\ \lambda^2+\cdots\end{equation}
where $v_g=\partial\Omega/\partial K$ is the group velocity and
$2P=\partial^2\Omega/\partial K^2$ the group velocity dispersion.
In that case, the expression for the wave packet becomes
\begin{equation}
u_{n}(t)=\epsilon e^{i(Kn-\Omega t)}\phi(Z,T),\end{equation}
where the slowly  varying envelope $\phi$, namely
\begin{equation}
\phi(Z,T)=\int d\lambda\ \hat u(\lambda)
e^{i(\lambda Z-\lambda^2P T)}\end{equation}
is written in the frame
\begin{equation}
Z=\epsilon(n-v_gt)\ ,\quad T=\epsilon^2t.\end{equation}
This means that the effects of an initial disturbance are observed for large
($\epsilon^{-2}$) times by following the waves at the group velocity and
analyzing the perturbation about the plane wave solution.  This approach then
corresponds to treating {\em initial value problems}. Note in particular that
the variable $Z$ hereabove defined becomes by construction a continuous 
variable.

Alternatively, considering the wave packet
\begin{equation}
u_{n}(t)=\int d\omega\ \hat u(\omega) e^{i(kn-\omega t)},
\end{equation}
we may expand here the wave number in terms of the frequency as
\begin{equation}
\omega=\Omega+\epsilon\nu\ ,\quad
k=K+\epsilon\frac{\nu }{v_g}+
\epsilon^2 Q\ \nu^2+\cdots\end{equation}
(with $2Q=\partial^2K/\partial \Omega^2$), hence giving
\begin{equation}
u_{n}(t)=\epsilon e^{i(Kn-\Omega t)}\psi(\xi,\tau),\end{equation}
with the slow modulation
\begin{equation}
\psi(\xi,\tau)=\int d\nu\ \hat u(\nu)e^{i(\nu^2Q\xi-\nu\tau)},\end{equation}
in the frame
\begin{equation}
\label{slow-var}\tau=\epsilon(t-n/v_g)\ ,\quad \xi=\epsilon^2n .
\end{equation}
Such a reference frame has a clear physical meaning: one considers long
distance ($\epsilon^{-2}$) effects in the retarded time to let the input
disturbance enough time to reach the observed lattice point.  Hence this
corresponds to the situation where the lattice is excited at one end, in other
words to a {\em boundary value problem}. 
This is the tool we shall use hereafter when the variable $\xi$ is kept discrete
as follows.

\paragraph*{Discrete multiscaling.}
Performing a discrete multiscale analysis goes first with assuming
a hypothesis of {\em slow modulation}. This means for the envelope
$\psi_n(t)$ of the wave $A(n,t)$, that
\begin{equation}\label{hyp-slow}
||\partial_t\psi_n(t)||\sim \epsilon||\partial_t A(n,t)||\ ,
\end{equation}
which then allows to chracterize the dimension $\epsilon$ of the 
slow variable $\tau$ by (\ref{slow-var}).

Second, as introduced in \cite{gros},  we define
a large grid indexed with the  slow variable $m$ defined  by sampling
the original grid at each  $N=\epsilon^{-2}$ (more precisely the closest 
odd integer). As a consequence we can index the variable $\xi_n$ by $m$ 
in the new grid, namely we can call $m$ a given point $n$ and $m+j$ the
points $n+jN$ for all $j$. In short
\begin{equation}\label{xi-n}
\cdots,\ \xi_{n-N}=m-1,\ \xi_n= m,\  \xi_{n+N}=m+1,\ \cdots\ .
\end{equation}

Considering now a {\em slow modulation} $\psi(\xi_n,\tau_n)$ in the variables
(\ref{slow-var}), of the plane wave $A(n,t)$, we represent this modulation by
the function $\psi(m,\tau)$ of the above defined discrete variable $m$ and of
$\tau=\tau_n$.
Then it is demonstrated in \cite{gros} that the first centered difference
of $u_n(t)=A(n,t)\psi(m,\tau)$ reads
\begin{eqnarray}\label{first-prod-deriv}
u_{n+1}&&-u_{n-1}=[A_{n+1}-A_{n-1}]\psi_{m}\nonumber\\
&&+(\epsilon/v_g)[A_{n+1}+A_{n-1}]\partial_\tau\psi_{m}\nonumber\\
&&+\epsilon^2/2[A_{n+1}+A_{n-1}][\psi_{m+1}-\psi_{m-1}]\nonumber\\
&&+(\epsilon/v_g)^2\frac12[A_{n+1}-A_{n-1}]\partial_\tau^2\psi_m 
+{\cal O}(\epsilon^3)\ ,
\end{eqnarray}
while the second discrete derivative is given by
\begin{eqnarray}\label{second-prod-deriv}
u_{n+1}&&-2u_{n}+u_{n-1}
=[A_{n+1}-2A_{n}+A_{n-1}]\ \psi_{m} \nonumber\\
&&+(\epsilon/v_g) [A_{n+1}-A_{n-1}]\partial_\tau \psi_{m} \nonumber\\
&&+(\epsilon/v_g)^2\frac12[A_{n+1}+A_{n-1}]\partial_\tau^2\psi_{m}\nonumber\\
&&+\epsilon^2/2[A_{n+1}-A_{n-1}] [\psi_{m+1}-\psi_{m-1}]+{\cal O}(\epsilon^3)\ .
\end{eqnarray}

\paragraph*{The discrete electrical lattice.}

We consider now the problem of the onset of modulational instability in the 
nonlinear electrical lattice  experimentaly and
theoreticaly studied in \cite{remsnet-elec-dis}. The model reads
\begin{eqnarray}\label{elec-model}
(A+V_n)\partial_t^2V_n-(\partial_tV_n)^2=\nonumber\\
\frac{u_0^2}A(A+V_n)^2\left[V_{n+1}+V_{n-1}-
(2+\frac{\omega_0^2}{u_0^2})V_n\right]\ ,\end{eqnarray}
where $A$ is a constant and $V_n(t)$ is the voltage of the $n$-th 
cell. The linearized version of this equation does possess the dispersion 
relation (\ref{lin-disp}). The experiments consist in sending at one end
of the electrical lattice a voltage $V_0(t)$ as a slightly modulated
plane wave for various values of the carrier frequency and amplitude, and
then measuring the output voltage at some cells along the lattice.

To proceed with discrete multiscaling, it is more
convenient to work with the equivalent system
\begin{eqnarray}\label{elec-sys}
&\partial_tV_n=u_0\left[1+\frac1AV_n\right]B_n\ ,\nonumber\\
&\partial_tB_n=u_0\left[V_{n+1}+V_{n-1}-
(2+\frac{\omega_0^2}{u_0^2})V_n\right]\ .\end{eqnarray}
A perturbation expansion is now performed with the following series
\begin{eqnarray}\label{series}
&B_n=\sum_{p=1}^{\infty}\epsilon^p\ \sum_{\ell=-p}^{p}
A^\ell(n,t)\psi_p^{(\ell)}(m,\tau)\ ,\nonumber\\
&V_n=\sum_{p=1}^{\infty}\epsilon^p\ \sum_{\ell=-p}^{p}
A^\ell(n,t)\phi_p^{(\ell)}(m,\tau)\ ,\end{eqnarray}
where of course $A^\ell(n,t)=\exp[i\ell(Kn-\Omega t)]$, and where the reality 
condition results in $\psi_p^{(-\ell)}=\bar\psi_p^{(\ell)}$ and
$\phi_p^{(-\ell)}=\bar\phi_p^{(\ell)}$.

Inserting the above expressions in (\ref{elec-sys}) and using the formula
(\ref{second-prod-deriv}), we arrive at
\begin{eqnarray}\label{relat}
&&\phi_1^{(0)}=\psi_1^{(0)}=0\ ,\quad\phi_1^{(1)}= \eta \ ,
\nonumber\\ 
&& \psi_1^{(1)}= -i\frac\Omega{u_0}\eta  \ ,\quad
\phi_2^{(0)}=\psi_2^{(0)}=0 \ ,\nonumber\\
&&  \phi_2^{(1)}=-\frac{i}{\Omega}[\partial_\tau\eta -u_0
   \delta ]
\ ,\quad \psi_2^{(1)}=\delta \ ,\nonumber\\
&&\phi_2^{(2)}=-2\frac{\Omega^2}{Aa}\eta ^2\ ,\quad
\psi_2^{(2)}=i\frac\Omega{u_0}[\frac{4\Omega^2}{Aa}+\frac1A]\eta ^2
\ ,\nonumber\\
&&\phi_3^{(0)}=0\ ,\quad\psi_3^{(0)}=-\frac1{Au_0}\partial_\tau|
\eta |^2 \ ,\end{eqnarray}
where $\eta=\eta_m(\tau)$ and $\delta=\delta_m(\tau)$ are the new unknowns and
where  we have defined
$$a=-3\omega_0^2-4u_0^2(1-\cos K)^2\ .$$
Solving then the system for $\phi_3^{(1)}$ and $\psi_3^{(1)}$, the unknown 
function  $\delta_m(\tau)$ cancels out and we are left with the equation
for $\eta_m(\tau)$ which eventually writes
\begin{equation}\label{LM}
\frac i2 [\eta_{m+1}-\eta_{m-1}]+
Q\ \frac{\partial^2\eta_m}{\partial \tau^2}
-\gamma|\eta_m|^2\eta_m=0\ ,\end{equation}
with the following  $K$-dependent coefficients
\begin{eqnarray}\label{coeffs}
&&Q=\frac12\frac{\partial^2K}{\partial\Omega^2}=
\frac{u_0^2(1-\cos K)^2-\omega_0^2\cos K}{2u_0^4\sin^3 K}\ ,\nonumber\\
&& \gamma=\frac1{2A^2}\ 
\frac{\Omega^2[\omega_0^2+4u_0^2\sin^2K]}
{u_0^2\sin K[3\omega_0^2+4u_0^2(1-\cos K)^2]}\ .
\end{eqnarray}

From (\ref{series}) and (\ref{relat}) we have then
\begin{equation}\label{eta-V}
V_n(t)=\epsilon\ \eta_m(\tau)e^{i(Kn-\Omega t)}+{\cal O}(\epsilon^2)+c.c.
\end{equation}
Note that higher order corrections to
$V(n,t)$ are either explicitely expressed in terms of $\eta$ through 
(\ref{relat}) or by linear inhomogeneous differential equations. That means
that the theory is self-consistent and in particular that the overtones are
by no means neglected.

\paragraph*{Modulational instability.}

The continuous version of the equation (\ref{LM}) is a well known model for 
boundary value problems in optical fibers \cite{hasegawa}. It has been studied
also in the case of initial value problems in unstable media \cite{wadati} to
describe for instance Rayleigh-Taylor instability.

The plane wave solution to  (\ref{LM}), say 
\begin{equation}\eta_m(\tau)=B\exp[i(\lambda m-\nu \tau)]\ ,\end{equation}
obeys the dispersion law
\begin{equation}\label{disp-pertu}
\nu^2=-\frac1Q [\sin\lambda+\gamma|B|^2]\ .
\end{equation}
As we are considering a boundary value problem, $\nu$ and $B$ are
the {\em given} frequency and amplitude of the modulation. Then, as 
$\sin(\lambda)$ is bounded, and this is precisely a discretreness effect
(see conclusion), there is no real solution $\lambda$ if
\begin{equation}\label{threshold}
Q>0\ ,\quad |B|^2>\frac{1}{\gamma}=B_{cr}^2\ ,\end{equation}
leading to unstable plane waves.

Reporting to the definition (\ref{coeffs})  we first remark that
$Q$  vanishes at $K=K_{cr}$ which is also the zero of 
the group velocity dispersion.
Second, it is easy to see that 
\begin{equation}\label{Kcr}
K<K_{cr}\Rightarrow Q<0\ ,\quad
K>K_{cr} \Rightarrow Q>0\ .\end{equation}
Then the waves are unstable for $K>K_{cr}$ and $|B|>|B_{cr}|$.

\paragraph*{Application.}

The parameters of the experiments of \cite{remsnet-elec-dis} are
\begin{equation}\label{param}
\omega_0=2.13\ {\rm MHz}\ ,\quad u_0=1.77\ {\rm MHz}\ ,\quad A=3.9\ {\rm V}\ ,
\end{equation}
for which the zero $K_{cr}$ of $\alpha(K)$ results to be
 \begin{equation}K_{cr}=1.25\ {\rm rad.cell}^{-1},\end{equation}
in agreement with the experimental observation of \cite{remsnet-elec-dis}.

Concerning the threshold amplitude $B_{cr}$ defined in (\ref{threshold}), its
actual value in the physical units is found by going from $\eta_m(\tau)$ to the
value of $V_n(t)$ through (\ref{eta-V}), namely the critical value of 
the amplitude of $V_n(t)$ is simply $\epsilon B_{cr}$. 
Hence we must evaluate $\epsilon$ by using its
very definition, the requirement of {\em slow} modulation (\ref{hyp-slow}).  In
one of the the experiments of \cite{remsnet-elec-dis} (see their fig. 6), an
input plane wave  of frequency $f=600\ Khz$ (wave number $K_e=2.15\
rad.cell^{-1}$) is modulated at $F=32\ Khz$ and sent in the lattice.
Remembering the change of variables (\ref{slow-var}), in particular
$\tau=\epsilon t$, such a modulation implies
$ \epsilon= f/F\sim 0.053$.

 For the parameters (\ref{param}), the critical amplitude $B_{cr}$, at wave
number $K_e$ is easily evaluated and hence value of the instability threshold
results to be 
\begin{equation}\epsilon B_{cr}(K_e)=0.23\ V\ .\end{equation}
This value is significantly smaller than those currently used in the
experiments of \cite{remsnet-elec-dis} where damping prevents the use of
smaller voltage input. However, the actual dependence of  $B_{cr}$ on
the relative frequencies of the carrier and modulation waves (through the
parameter $\epsilon$) allows one to vary the value of threshold $B_{cr}$.
This property could be used to check the threshold amplitude on new
experiments. As shown on the figure, we note finally that the function
$B_{cr}(K)$ is bounded (maximum of $4.68\ V$ at $K=2.59$), hence this
threshold of instability exists at any carrier wave number in $]K_{cr},\pi[$.
\begin{figure}[h]\begin{center}
\epsfig{file=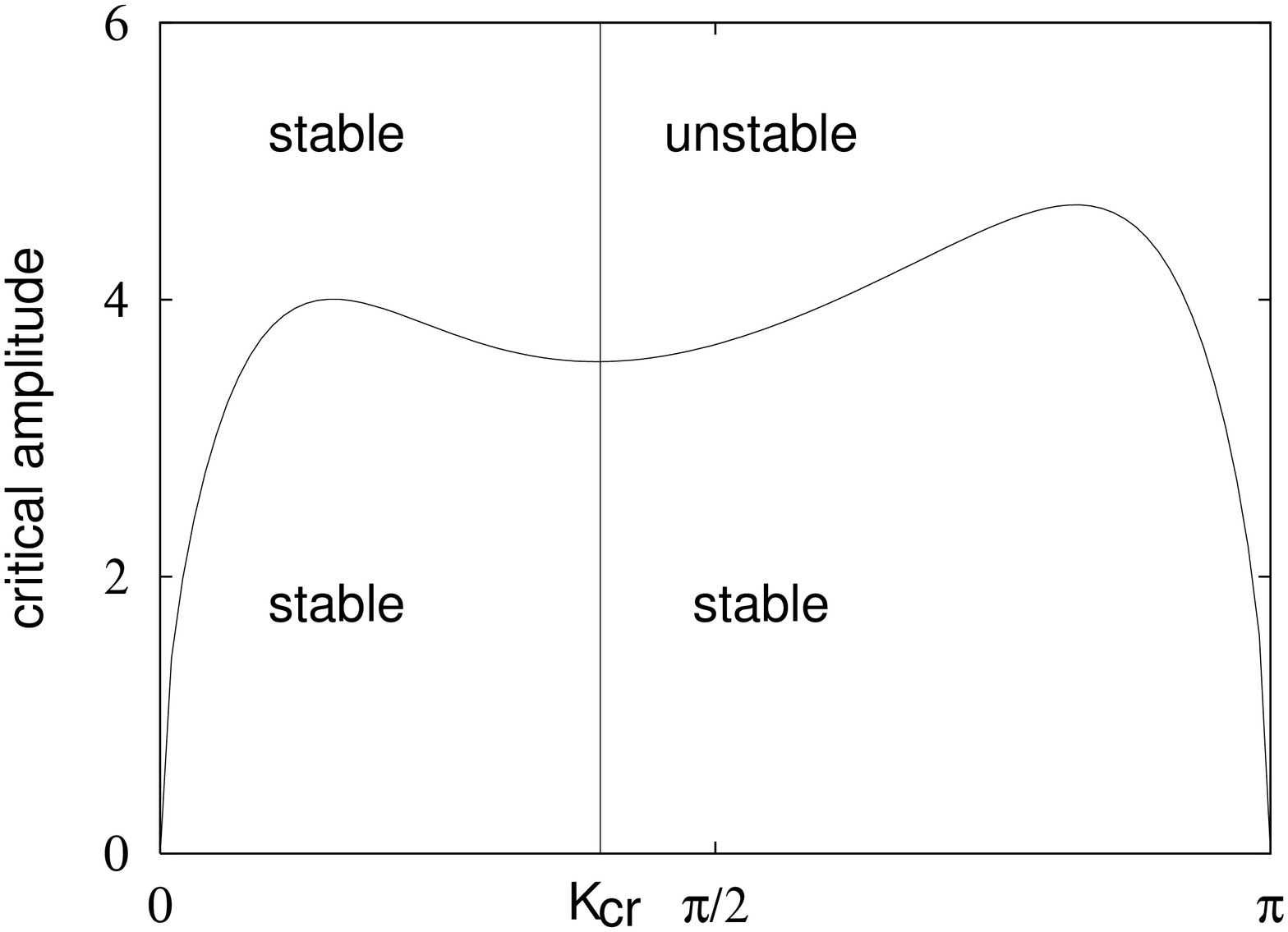,height=8cm,width=5cm,angle=-90}
\end{center}\end{figure}
\centerline{\em Fig.1:  
Critical amplitude $B_{cr}(K)$ in units of $\epsilon$ }   %

\paragraph*{Predictions.} 
The theory can also be applied to the sine-Gordon chain
\begin{equation}\label{SG}
\ddot u_n - u_0^2\left(u_{n+1}-2u_n+u_{n-1}\right)+\omega_0^2\sin u_n
=0\  ,\end{equation}
which has the linear dispersion relation (\ref{lin-disp}) and
for which the limit model is (\ref{LM}) with the coefficient
\begin{equation}\label{coeffs-sg}
\gamma=\frac{\omega_0^2}{4u_0^2\sin K}\ ,
\end{equation}
and of course the same value for $Q$.
Performing the analysis shows an instability for wave numbers greater than
the zero $K_{cr}$ of $Q(K)$,  which develops for modulation amplitude obeying
(in units of $\epsilon$ and in the Brillouin zone $K\in]0,\pi[$) 
\begin{equation}
B>B_{cr}=\frac{\omega_0}{2u_0\sqrt{\sin K}}.\end{equation}

For the integrable Toda lattice whose dispersion relation is
$\Omega=2\sin(K/2)$, leading to the limit model (\ref{LM}) with \cite{gros}
\begin{equation}
Q=-\frac14\frac{\sin(K/2)}{\cos^3(K/2)}\ ,\quad
\gamma=-\frac1{\sin K}\ ,\end{equation}
the theory implies that the waves are unstable for all $K$ and
$|B|^2>|\sin K|$.

\paragraph*{Conclusion.} 

The instability described above results from wave scattering in a nonlinear
medium (boundary value problem) which requires to use a change of variables
like (\ref{slow-var}) which in turn leads to the unstable wave equation
(\ref{LM}) for the envelope. The resulting instability is then of {\em first
order} (a second perturbation analysis of the equation for the envelope is not
required), and has for boundary value problems the same universal character as
the Benjamin-Feir instability has for initial value problems.

Moreover this instability is a purely discrete phenomenon as indeed,
performing a semi-discrete (discrete carrier and continuous envelope) analysis
with the new variables (\ref{slow-var}) leads to the continous version of
(\ref{LM}) and hence to the dispersion law
\begin{equation}\label{disp-cont}
\nu^2=-\frac1Q [\lambda+\gamma|B|^2]\ .
\end{equation}
The point is that, even for positive values of $Q$, there always exists a
real solution $\lambda$ of the above dispersion relation for any given
modulation frequency $\nu$.

\end{multicols}

\end{document}